\documentclass[12pt,letterpaper]{article}
\usepackage[T1]{fontenc}
\usepackage[utf8]{inputenc}
\usepackage{lmodern}
\usepackage{amsmath,amssymb,amsthm}
\usepackage{graphicx,booktabs,array,tabularx,longtable}
\usepackage[round,authoryear]{natbib}
\usepackage{url,xurl}
\usepackage[hidelinks]{hyperref}
\usepackage{tikz}
\usetikzlibrary{arrows.meta,positioning}
\usepackage[letterpaper,margin=1in,footskip=0.4in]{geometry}
\hypersetup{pdftitle={Causal Discovery beyond Conditional Independence},pdfauthor={Yang Ni},pdfsubject={Integrated review of causal discovery},pdfkeywords={causal discovery, DAG, feedback, Bayesian networks, heterogeneity, functional data}}
\newcommand{\spacingset}[1]{\renewcommand{\baselinestretch}{#1}\small\normalsize}
\newcommand{\pa}{\operatorname{pa}}
\newcommand{\Var}{\operatorname{Var}}
\newcommand{\Cov}{\operatorname{Cov}}
\newcommand{\E}{\mathbb{E}}
\newcommand{\indep}{\mathrel{\perp\!\!\!\perp}}
\newcommand{\R}{\mathbb{R}}
\newcommand{\G}{\mathcal{G}}

\newcommand{\D}{\mathcal{D}}
\theoremstyle{definition}
\newtheorem{example}{Example}[section]
\newcolumntype{Y}{>{\raggedright\arraybackslash}X}
\begin{document}
\spacingset{1.25}
\title{Causal Discovery beyond Conditional Independence:\\Identifiability, Algorithms, and Statistical Inference for Directed Acyclic and Cyclic Graphs}
\author{Yang Ni\\[3pt]
\small Department of Statistics and Data Science\\
\small The University of Texas at Austin\\
\small \texttt{yangni87@gmail.com}}
\date{}
\maketitle
\begin{abstract}
Learning causal structure from data requires assumptions about mechanisms, noises, observation schemes, and interventions. Historically, causal discovery has relied primarily on conditional independence. This review focuses on modern causal discovery methods that exploit properties of distributions beyond conditional independence. Our organizing principle separates the source of identification from the statistical estimator and its computational implementation. This perspective connects conditional-independence methods with distributional asymmetries, conditional-moment restrictions, and variation across environments or individuals. We examine discovery under latent confounding and feedback, distinguishing causal information about observed variables from reconstruction of latent structure and emphasizing the additional semantics required by cyclic models. We also distinguish population identification, frequentist and Bayesian consistency, uncertainty quantification, and algorithmic guarantees. Throughout, examples show why the same observational distribution can support different conclusions under different model classes. We conclude with guidance for comparing methods at their identifiable targets and a research agenda for uncertainty, heterogeneous populations, non-scalar variables, and the intersection of discrete data, feedback, and hidden confounding.
\end{abstract}
\noindent\textit{Keywords:} Bayesian networks; Feedback; Heterogeneity; Latent confounding; Structural equation models

\section{Introduction and historical perspective}
\label{sec:intro}

The same joint distribution can admit different causal explanations: an association may arise from a direct effect, a hidden common cause, or feedback at equilibrium. Increasing the sample size improves our knowledge of that distribution, but it cannot by itself distinguish explanations that generate the same observational distribution. Causal discovery therefore depends on assumptions that connect features of the observed distribution to causal structure. This review examines those connections and the statistical procedures used to exploit them.

Causal discovery finds many applications, with different measurements and opportunities for intervention. In systems biology, \citet{sachs2005} combine single-cell measurements with molecular perturbations to infer protein-signaling relationships, while count-based methods seek gene-regulatory structure from observational single-cell data \citep{choi2025}. In psychology, \citet{ni2025survey} use ordinal Bayesian networks to study relationships among obsessive-compulsive and depressive symptoms measured by questionnaires. In finance, \citet{jiang2023markets} apply VAR-LiNGAM and a latent-variable discovery method to investigate links among foreign-exchange, stock, and bond markets in Japan and the United States. These examples motivate several distinctions developed below: experimental versus observational data, continuous versus categorical or count measurements, and contemporaneous relationships versus temporal dependence.

Modern causal discovery draws on several historic developments. Path analysis represented causal hypotheses by systems of equations and decompositions of association long before automated graph search became practical \citep{wright1921}. Probabilistic graphical models---most prominently Bayesian networks, which are probabilistic models Markov with respect to directed acyclic graphs (DAGs)---subsequently provided a language for relating graph separation to conditional independence. The characterization of observational equivalence by \citet{verma1990} and the search framework of \citet{spirtes2000} established that useful causal information can sometimes be recovered even when a complete graph cannot. In parallel, Bayesian network learning developed scores that combine fit, complexity, and prior information \citep{heckerman1995}. Orientation rules and equivalence-class search made this framework algorithmically concrete \citep{meek1995,chickering2002}. Structural causal models supplied a common interpretation of interventions and counterfactuals \citep{pearl2009}.

A major development was the recognition that conditional independence does not exhaust the information in an observational distribution. The linear non-Gaussian acyclic model (LiNGAM) of \citet{shimizu2006} demonstrated that restrictions on noises can identify a complete causal DAG from observational data. The graph may be distinguishable even when its conditional-independence implications are identical to those of another graph. Later developments exploited nonlinear additive noise \citep{hoyer2008,peters2014}, conditional moments \citep{park2015,park2020}, restrictions on categorical mechanisms \citep{ni2022ordinal,ni2022colp}, and systematic variation across environments or experimental units \citep{peters2016,zhou2022hetero}. These results changed the question from which arrows are compelled by conditional independence to which assumptions generate a directional asymmetry in the distribution.

The usual division into constraint-based, score-based, and functional-model methods is useful for describing algorithms, but it combines two different questions: what identifies the causal structure, and how that information is extracted from data. For example, the same non-Gaussian or additive-noise model can be fitted by discrete graph search, continuous optimization, or Bayesian sampling. Conversely, a likelihood score may distinguish only an equivalence class under one model and a unique DAG under another. We therefore organize the review around sources of identifying information instead of learning algorithms, using the computational methods to show how each source can be put to work.

We review observational and interventional causal discovery from the LiNGAM era onward, with a short historical foundation and primary emphasis on observational identification. We cover acyclic and cyclic systems, both with and without latent confounding, and distinguish continuous, discrete, count, mixed, and functional variables. Interventions and heterogeneous environments enter as additional sources of information. We also include heterogeneity at the level of individuals: the aim may be to learn a family of causal structures rather than one graph shared by a population. We discuss temporal models when they clarify feedback or functional observations, but do not attempt a comprehensive review of time-series causal discovery.

Complementary reviews cover graphical and structural approaches \citep{spirteszhang2016} and statistical structure learning \citep{drton2017}. Here the emphasis is on connections among identification, estimation, computation, and uncertainty across graph and data types.
We prioritize papers that establish a distinct identifying principle, a statistical guarantee, or a computational approach that changes how a model class can be fitted.

\section{A common language for models and inferential targets}
\label{sec:framework}

\subsection{Structural equations, observations, and interventions}

Let $X=(X_1,\ldots,X_p)$ denote the random variables of interest. Let $\G=(V,E)$ denote a directed graph with a set of vertices or nodes $V=\{1,\dots,p\}$ representing $X$ and a set of directed edges $E$ representing direct causal relationships among the nodes. We say $k$ is a parent of $j$ if $k\to j\in E$ and collect all the parents of $j$ in $\G$ as $\pa_\G(j)$. A structural causal model specifies an assignment to each variable $X_j$ through its parents $X_{\pa_\G(j)}$, an exogenous noise variable $U_j$, and a deterministic function $f_j$, for $j=1,\ldots,p$,
\begin{equation}
 X_j=f_j(X_{\pa_\G(j)},U_j).
 \label{eq:scm}
\end{equation}
We initially assume that the graph is acyclic and that the noises are mutually independent. These assumptions give a standard causally sufficient model, with no unmeasured common causes, in which variables can be generated recursively following a causal order. Relaxing either assumption changes this construction. Marginalizing unmeasured confounders can induce dependence among the observed-variable noises, while allowing feedback removes the recursive ordering. In the latter case, the structural equations must be supplemented by conditions that ensure existence and uniqueness of their solution \citep{bongers2021}.

For an acyclic model with independent noises, the observational distribution admits a convenient factorization, written in terms of conditional densities or probability mass functions when they exist:
\begin{equation*}
 p(x)=\prod_{j=1}^p p_j(x_j\mid x_{\pa_\G(j)}).
\end{equation*}
An intervention replaces one or more assignments in \eqref{eq:scm}. A perfect intervention fixes a variable or replaces its mechanism independently of its original parents. A soft intervention may alter a mechanism while retaining some parental dependence. 

\subsection{Equivalence and identification}

Two DAGs are Markov equivalent when they imply the same set of conditional independences. For DAGs, equality of skeletons and unshielded colliders (also known as v-structures) characterizes this equivalence \citep{verma1990}. A completed partially directed acyclic graph (CPDAG) represents a Markov equivalence class: its directed edges have the same orientation in every member, whereas undirected edges remain reversible within the class.

To recover structure from conditional independences, one usually combines the Markov property with faithfulness. The Markov property supplies the independences implied by graph separation, whereas faithfulness rules out additional independences caused, for example, by cancellation of effects. Finite-sample recovery requires more than ruling out exact cancellation: uniform or high-dimensional guarantees often need quantitative bounds that keep relevant associations away from zero. These requirements belong to conditional-independence approaches; other identifying arguments can use different information. LiNGAM, for instance, uses independent non-Gaussian noises and independent component analysis to identify a DAG.

A broader definition makes the role of a model class explicit. Let $\mathcal M$ be a collection of structural models, let $P_M^{\mathrm{obs}}$ be the observed law under $M\in \mathcal M$, and let $\tau(M)$ be the causal target. At a model $M_0$, the target is identified if
\begin{equation*}
 \{\tau(M):M\in\mathcal M,\ P_M^{\mathrm{obs}}=P_{M_0}^{\mathrm{obs}}\}
 =\{\tau(M_0)\}.
\end{equation*}
The target may be a graph, an equivalence class, a causal order, an ancestral relation, or a causal effect.

Which structures remain equivalent depends on the statistical model. Two DAGs can imply the same conditional independences while generating different sets of distributions once non-Gaussian noises or nonlinear mechanisms are imposed. Those additional restrictions can therefore orient edges left undirected by a CPDAG. If Gaussian and non-Gaussian noises coexist, however, they may resolve only part of the graph. The appropriate target is then a distribution-equivalence class that is finer than a Markov equivalence class but still contains several DAGs. Section~\ref{sec:bayesian-selection} returns to this intermediate case and its consequences for inference.

\begin{example}
\label{ex:run}
A running example will connect the identifying restrictions to a common target. Consider three centered variables with independent noises,
\begin{equation}
 X=\varepsilon_X,\qquad Y=bX+\varepsilon_Y,\qquad
 W=cY+\varepsilon_W,\qquad b,c\ne0.
 \label{eq:running-chain}
\end{equation}
The underlying DAG is $X\to Y\to W$.
The target is the total effect $\tau=\partial\E\{W\mid\operatorname{do}(X=x)\}/\partial x=bc$. With Gaussian errors, the same faithful observational law admits three DAGs: $X\to Y\to W$, $X\leftarrow Y\to W$, and $X\leftarrow Y\leftarrow W$. All imply $X\indep W\mid Y$, but the latter two imply $\tau=0$. Thus the CPDAG $X-Y-W$ determines neither direction nor effect. Sections~\ref{sec:observational}, \ref{sec:latent}, \ref{sec:cycles}, and~\ref{sec:heterogeneity} revisit this system as noise restrictions, hidden causes, feedback, and interventions change the information available. Section~\ref{sec:uq} returns to uncertainty about $\tau$.
\end{example}
\subsection{Four layers of a discovery procedure}

We use four connected questions to organize the remainder of the review. First, which causal features does the population distribution identify under the stated model? Once that target is clear, the second question is how to estimate it, for example through tests or a penalized likelihood. The third concerns computation: an implemented search or approximation must calculate, or approximate, the proposed estimator. Finally, an analysis needs an assessment of uncertainty and of its sensitivity to the assumptions that made identification possible. Keeping these questions separate helps explain why a computational improvement need not strengthen identification, and why a consistent estimator need not come with a calibrated uncertainty statement.

\section{Observational DAGs: information beyond conditional independence}
\label{sec:observational}

\subsection{Conditional-independence and equivalence-class discovery}

The classic PC algorithm first searches for a skeleton by testing conditional independences, identifies unshielded colliders, and propagates orientations. Its output targets a CPDAG under the causal sufficiency, Markov, and faithfulness assumptions \citep{spirtes2000}. Sparsity can make the procedure attractive because it limits the size of candidate conditioning sets. \citet{kalisch2007} established consistency in high-dimensional sparse Gaussian settings under additional regularity and signal conditions. \citet{colombo2014} showed that the order in which variables enter the procedure can change the learned graph in a finite sample, and order-independent modifications improve reproducibility.

The conditional-independence test determines which data distributions a PC analysis can accommodate. Partial-correlation tests are appropriate under joint Gaussianity, contingency-table tests handle categorical variables, and kernel methods provide tests for continuous nonlinear relationships \citep{zhang2011}. Thus, although the graphical search can remain the same, its choice of tests changes with the variable type and model assumptions. Section~\ref{sec:theory} examines how the validity and power of these tests affect guarantees for the resulting graph.

Score-based procedures approach the same structure learning question by comparing a likelihood, complexity-penalized fit, or marginal likelihood across candidate graphs. Greedy equivalence search (GES), for example, searches over equivalence classes and has asymptotic guarantees under appropriate scoring and distributional assumptions \citep{chickering2002}. Hybrid procedures combine this search with conditional-independence tests: max-min hill climbing first uses such information to restrict the candidate edges \citep{tsamardinos2006}. The restriction can save substantial computation, but a mistaken exclusion may prevent the subsequent search from reaching the target. Adaptively restricted GES addresses this difficulty by allowing the search space to change during learning; \citet{nandy2018} establish high-dimensional consistency for GES and its adaptive restriction under stated conditions.

An alternative is to search over causal orders. Each permutation determines which variables are eligible parents of a given node, and hence graph construction within that order automatically respects acyclicity. The sparsest permutation principle selects a sparsest compatible graph and can be justified under assumptions weaker than faithfulness \citep{raskutti2018}. 

For a complementary review of equivalence classes and combinatorial search, see \citet{squires2023}.

These approaches provide the background for the distributional methods below: they use conditional independences or model scores to restrict the possible structures, while additional restrictions on mechanisms and noises can distinguish directions within an equivalence class.

\subsection{LiNGAM and independent component analysis}
The equivalence-class targets above reflect how much information conditional independences can supply. LiNGAM \citep{shimizu2006} showed how restrictions on the full distribution can distinguish directions within such a class. To see the additional information it uses, write a linear structural causal model as
\begin{equation}
 X=BX+\varepsilon,\qquad X=(I-B)^{-1}\varepsilon,
 \label{eq:lingam}
\end{equation}
where $B_{jk}$ is the direct effect of $X_k$ on $X_j$ and $B_{jj}=0$. Under acyclicity, $B$ can be permuted to a strictly triangular matrix. LiNGAM assumes independent non-Gaussian noise and uses independent component analysis to identify the mixing matrix $(I-B)^{-1}$ in \eqref{eq:lingam} up to its familiar scaling and permutation ambiguities. The structural normalization and acyclic ordering resolve these ambiguities.

DirectLiNGAM \citep{shimizu2011} exploits the same identifying information differently. A candidate exogenous variable is compared with residuals obtained by regressing other variables on it. Under the model, a true exogenous variable is independent of the relevant residuals, enabling recursive extraction of an order. This connects ICA-based identification with regression and independence testing. Pairwise likelihood ratios provide another computational route \citep{hyvarinen2013}, and high-dimensional extensions use sparsity and moment-based information to control ordering errors \citep{wang2020}.

Revisiting the example in \eqref{eq:running-chain} makes the residual asymmetry explicit. Write $v_X=\Var(\varepsilon_X)>0$ and $v_Y=\Var(\varepsilon_Y)>0$. The population reverse least-squares coefficient and residual (regressing $X$ on $Y$) are
\begin{equation*}
 a=\frac{b v_X}{b^2v_X+v_Y},\qquad
 R_{X\mid Y}=X-aY=(1-ab)\varepsilon_X-a\varepsilon_Y.
\end{equation*}
Although $\Cov(Y,R_{X\mid Y})=0$, both $Y$ and $R_{X\mid Y}$ contain both noises $\varepsilon_X$ and $\varepsilon_Y$ when $b\ne0$. The Darmois--Skitovich theorem states that if two linear forms of mutually independent random variables are independent, every variable entering both forms with a nonzero coefficient must be Gaussian \citep{skitovich1954}. Independence of these forms would therefore require both noises to be Gaussian. In contrast, the residual under the correct model is independent of the cause, $\varepsilon_Y\indep X$. This calculation explains why testing correlation cannot implement LiNGAM: least-squares residuals are uncorrelated with the predictor, whereas independence resolves the causal direction. DirectLiNGAM generalizes the residual comparison to selecting exogenous variables and recursively removing their contributions.

With independent non-Gaussian errors throughout \eqref{eq:running-chain}, LiNGAM identifies both arrows and hence the effect $\tau=bc$: the chain and fork no longer represent the same observational law within this model class. The assumptions also indicate where this argument can fail. Gaussian noises can leave some directions unresolved because the required independence asymmetry disappears. Hidden common causes and measurement error create a different challenge by introducing additional latent components into the observed mixtures. Even when the assumptions hold exactly, noises close to Gaussian can supply little directional information in a finite sample. These issues motivate extensions that either model the additional components or seek other sources of identification.

\subsection{Variance restrictions as an alternative source of orientation}

Non-Gaussianity is sufficient but not necessary for observational identification. \citet{petersbuhlmann2014} show that a linear Gaussian DAG with equal error variances is identifiable. In a causally sufficient linear model with common noise variance $\sigma^2$, a root has variance $\sigma^2$, whereas a nonroot has variance equal to $\sigma^2$ plus the variance of its parental contribution, provided that contribution is nondegenerate. Conditional versions of this reasoning enable recursive ordering. In \eqref{eq:running-chain}, equal error variances give $\Var(X)=\sigma^2$, $\Var(Y)=(1+b^2)\sigma^2$, and $\Var(W)=\{1+c^2(1+b^2)\}\sigma^2$. The minimum identifies $X$ as a root; after removing its contribution, the same argument orders $Y$ before $W$. This recovers $\tau=bc$ even with Gaussian noises, using a restriction on scale instead of distributional shape.

\citet{loh2014} connect variance-weighted scoring with inverse-covariance estimation, and \citet{park2020} develop broader conditional-variance conditions that allow unequal noise variances and nonlinear assignments. \citet{dallakyan2025} generalize variance-based criteria using weak majorization: the noise-variance vector is compared with the vector of expected conditional variances obtained along a candidate ordering. Weak majorization compares cumulative sums of the largest entries, allowing a vector-level restriction rather than requiring each variance comparison separately. Their sufficient criterion encompasses earlier conditional-variance conditions. For linear models, squared Cholesky diagonal entries are variances of successive linear-projection residuals; along a causal ordering they equal the structural error variances. Under arbitrary non-Gaussian orderings they need not equal expected conditional variances. Their MaCho algorithm uses this covariance-based representation. Together these results suggest a general strategy: identify an ordering through a population contrast whose extremum distinguishes a root or a sink, and then estimate parent sets given that ordering.

The interpretation of a variance restriction depends on the units in which variables are measured. Rescaling one variable changes its structural noise variance without changing the causal graph, and hence it can destroy an equal-variance assumption that was meaningful in the original units. Consequently, standardization is not a harmless preprocessing step for these methods and should be treated as a modeling choice. 

\subsection{Nonlinear additive noise and post-nonlinear mechanisms}

The additive-noise model takes the form
\begin{equation}
 X_j=f_j(X_{\pa_\G(j)})+\varepsilon_j,\qquad
 \varepsilon_1,\ldots,\varepsilon_p\ \text{mutually independent}.
 \label{eq:anm}
\end{equation}
\citet{hoyer2008} demonstrate that nonlinearities can make the forward model distinguishable from its reverse. Replacing $bX$ and $cY$ in \eqref{eq:running-chain} by nonlinear mechanisms connects this example to identifiable additive-noise subclasses; the intervention mean is then generally a nonlinear function of $x$, rather than a constant slope $bc$. A reverse conditional distribution generally cannot be expressed as a translated copy of one common noise distribution.
Specifically, for a pair of variables, an additive-noise fitting procedure therefore compares
\begin{equation*}
 D_{X\to Y}=\mathsf{Dep}\{X,Y-f(X)\},\qquad
 D_{Y\to X}=\mathsf{Dep}\{Y,X-g(Y)\},
\end{equation*}
where $\mathsf{Dep}$ is a dependence measure, which vanishes exactly at independence within the distributions under consideration. A suitable kernel dependence measure is one choice. Under an identifiable forward model, the correct $f$ gives $D_{X\to Y}=0$, while no admissible reverse function gives zero. Regression estimates replace $f$ and $g$ in samples, after which residual independence is assessed on held-out observations or with suitable calibration.

\citet{peters2014} extend additive-noise identification to multivariate systems by imposing conditions on the causal mechanisms and noise distributions. In their regression with subsequent independence testing (RESIT), a sink can be regressed on all remaining variables: its noise is independent of those variables, since none is its descendant. With population regressions and an independence oracle, recursive sink removal and subsequent edge pruning recover the graph under their restricted-model and causal-minimality conditions.

Once nonlinear mechanisms are allowed, fitting each node on many parents can become a difficult regression problem. The causal additive model (CAM) addresses this by imposing the further decomposition $f_j(x_{\pa(j)})=\sum_{k\in\pa(j)}f_{jk}(x_k)$, which facilitates order search and penalized regression \citep{buhlmann2014}. This, however, excludes interactions that a general additive-noise model can accommodate. When interactions are scientifically plausible, tree-based additive-noise DAGs offer an interpretable, more flexible approach \citep{zhou2025tree}.

Other extensions relax the way noise enters a mechanism. Post-nonlinear models permit an invertible transformation after noise is added, as in $Y=g(f(X)+\varepsilon)$; \citet{zhang2009} characterize circumstances in which this richer class remains directionally identifiable. A different extension allows the noise magnitude itself to depend on the cause. In a location-scale model, $Y=f(X)+s(X)\varepsilon$ with $s(X)>0$, both the conditional mean and conditional variance can change with $X$. \citet{immer2023} establish identification and estimation results for this setting. Together, these developments show how restrictions beyond homoscedastic additive noise can create an asymmetry between the two directions.

These procedures use regression as a means of testing a causal restriction. Good prediction alone does not select a direction: a reverse regression may predict accurately while leaving dependent residuals. Conversely, an inadequately fitted forward mechanism can produce residual dependence even when its causal direction is correct. An analysis must therefore assess regression fit together with the independence restriction. 

\section{Computation across identifying model classes}
\label{sec:computation}

\subsection{Search spaces and statistical objectives}

Once an identifying model has been chosen, computation requires both a statistical objective and a space over which to search. Searching directly over DAGs makes graph constraints explicit. Equivalence-class search avoids separately considering orientations that receive the same score, whereas order search enforces acyclicity through a permutation, at the cost of potentially representing the same DAG by several orders. Restrictions based on preliminary tests can reduce any of these searches, but they also allow early statistical errors to constrain later choices.

A generic score-based estimator can be written as
\begin{equation*}
 (\widehat\G,\widehat\theta)\in
 \arg\min_{\G\in\D,\,\theta\in\Theta_\G}
 \{-\ell_n(\theta;\G)+\lambda_n\operatorname{pen}(\G,\theta)\},
\end{equation*}
where $\D$ is the DAG space, $\ell_n$ is a log-likelihood or another fit criterion, and $\operatorname{pen}$ penalizes complexity. The model determines which structures the fit criterion can distinguish; for example, an unrestricted Gaussian likelihood does not distinguish Markov-equivalent DAGs. The penalty then affects which representation is selected among competing fits. Even after these statistical choices are justified, a computational question remains: guarantees for a global optimum, such as those of \citet{aragam2019}, apply to the estimator defined by the optimization problem, and an algorithm must calculate that estimator to inherit those guarantees.

DAG searching is challenging. The number of labeled DAGs grows faster than exponentially in the number of nodes; even order search has $p!$ possible permutations. An edge change must preserve acyclicity, and a locally favorable move can prevent a better configuration from being reached. Exact optimization can certify a global solution for manageable problems, whereas greedy search trades that guarantee for speed. These difficulties motivate differentiable representations of the constraint, which we review next.

\subsection{Continuous formulations of acyclicity}

NOTEARS \citep{zheng2018} introduced an exact smooth characterization of acyclicity for a weighted adjacency matrix:
\begin{equation}
 h(B)=\operatorname{tr}\{\exp(B\circ B)\}-p=0,
 \label{eq:acyclicity}
\end{equation}
where $\circ$ denotes entrywise multiplication. Squaring entries makes the contributions of directed walks nonnegative so that cycles cannot disappear through cancellation between positive and negative edge weights. The equality in \eqref{eq:acyclicity} therefore permits acyclicity to be imposed through a differentiable constraint. This is a computational reformulation of the graph restriction; the resulting optimization remains nonconvex, and identification still depends on the statistical model. Moreover, the numerical tolerance used to enforce the equality and the subsequent thresholding of edge weights affect whether the reported graph is exactly acyclic.

The original NOTEARS formulation fits a linear DAG by penalized least squares. For observations stored as rows of $\mathbf X\in\R^{n\times p}$, and with $B_{jk}$ representing $k\to j$, its optimization problem is
\begin{equation*}
 \min_{B:\,B_{jj}=0}\
 \frac{1}{2n}\|\mathbf X-\mathbf X B^{\mathsf T}\|_F^2
       +\lambda\|B\|_1
 \quad\text{subject to }h(B)=0.
\end{equation*}
Here $\|B\|_1$ sums absolute entries and $\|\cdot\|_F$ is the Frobenius norm. The acyclicity argument follows by expanding the matrix exponential as $h(B)=\sum_{m\ge1}\operatorname{tr}\{(B\circ B)^m\}/m!$. Each trace sums the weights of closed directed walks of length $m$, and all of these weights are nonnegative. A directed cycle therefore contributes a positive term, whereas a DAG has no closed directed walks. To enforce the equality numerically, an augmented-Lagrangian implementation adds $\alpha h(B)+\eta h(B)^2/2$ to the objective and updates the multiplier $\alpha$ and penalty $\eta>0$ during optimization.

The same strategy can be extended beyond linear models. DAG-GNN uses a neural generative model \citep{yu2019}, while GraN-DAG represents the conditional mechanisms with neural networks \citep{lachapelle2020}. Nonparametric NOTEARS instead links an edge to the dependence of a mechanism on an input through functional derivatives \citep{zheng2020}. In each case, the representation connects a flexible mechanism to a differentiable acyclicity restriction.

Other work changes the objective or the acyclicity characterization. GOLEM investigates the roles of likelihood, sparsity, and soft DAG constraints in linear models \citep{ng2020}. DAGMA replaces the matrix exponential with a log-determinant characterization \citep{bello2022}. Let $\rho(A)$ denote the spectral radius of a matrix, the largest absolute value of its eigenvalues. DAGMA chooses $s>\rho(B\circ B)$ and uses
\begin{equation*}
 h_s(B)=-\log\det\{sI-B\circ B\}+p\log s,
 \qquad s>\rho(B\circ B).
\end{equation*}
The condition $s>\rho(B\circ B)$ makes $sI-B\circ B$ invertible with positive determinant so that the logarithm is well defined. On this domain, $h_s(B)=\sum_{m\ge1}\operatorname{tr}\{(B\circ B)^m\}/(m s^m)$ is nonnegative and equals zero exactly for a DAG. Thus both characterizations detect cycles through closed walks, but weight them differently. DAGMA follows a sequence of penalized optimization problems, keeping the iterates in this domain while increasing the relative influence of the acyclicity term.

MARCEDES \citep{chaudhuri2026marcedes} connects non-Gaussian identification to continuous optimization through a profiled working Laplace likelihood. For zero-diagonal $B$ with $I-B$ invertible, its mean absolute residual risk is
\begin{equation*}
 L_n(B)=-\log|\det(I-B)|+
 \sum_{j=1}^{p}\log\left\{\frac1n\sum_{i=1}^{n}
 \left|X_j^{(i)}-\sum_k B_{jk}X_k^{(i)}\right|\right\}.
\end{equation*}
The first term accounts for the transformation from observations to residuals, while the remaining terms measure residual magnitudes after profiling the Laplace scales. To encourage sparse acyclic solutions, the fitting criterion adds equation-specific sparsity penalties and a soft acyclicity penalty. A variational surrogate and generalized-Bayes tuning lead to an alternating optimization scheme. The paper's pairwise consistency theorem compares the risk at the truth with that at a fixed competing matrix under independent Gaussian-scale-mixture errors with nondegenerate mixing distributions and finite first moments. This is a pairwise statistical argument, distinct from a guarantee for the computed optimizer. Although the determinant is the same type of change-of-variables term that appears in cyclic likelihoods in Section~\ref{sec:cycles}, MARCEDES uses it here to estimate an acyclic structure.

\subsection{Ordering from distributional scores}

Continuous optimization is not the only way to avoid a direct search over all graphs. Another approach extracts a causal ordering from derivatives of the observed density. Consider the score function $s(x)=\nabla_x\log p(x)$ (not to be confused with scores assigned to candidate graphs for score-based methods). For additive Gaussian-noise models, its derivatives have a distinctive form at sinks of the DAG. SCORE \citep{rolland2022} uses that property to recover an ordering under nonlinear identifiability conditions, and subsequent work \citep{montagna2023} extracts further graph information while reducing the cost of edge pruning.

For independent $\varepsilon_j\sim N(0,\sigma_j^2)$ in \eqref{eq:anm}, let $r_j(x)=x_j-f_j(x_{\pa(j)})$. Differentiating the factorized log density yields
\begin{equation*}
 s_j(x)=-\frac{r_j(x)}{\sigma_j^2}
   +\sum_{k:\,j\in\pa(k)}\frac{r_k(x)}{\sigma_k^2}
                       \frac{\partial f_k}{\partial x_j}.
\end{equation*}
If $j$ is a sink, it has no children and the sum vanishes. Differentiating once more with respect to $x_j$ then gives the constant $\partial s_j(x)/\partial x_j=-1/\sigma_j^2$, whose variance across observations is zero. SCORE estimates this derivative and searches for nodes with small estimated variation, relying on conditions that make the derivative non-constant at non-sinks \citep{rolland2022}. The role of nonlinearity is evident from the failure case: with entirely linear mechanisms, the joint density is Gaussian and its Hessian is constant at every node. The derivative calculation then supplies no distinction between sinks and non-sinks.

In a sample, this argument requires accurate estimates of density derivatives. Errors can also accumulate because each selected sink is removed before the next is considered, making control of successive ordering steps particularly important in high dimensions. For discrete data, ordinary density derivatives are unavailable, but a related strategy can be built from an appropriate discrete score operator; Section~\ref{sec:discrete} discusses the connection developed by \citet{vo2026}.

\subsection{Bayesian graph learning}
Bayesian graph learning targets a posterior over structures and parameters. In structure Markov chain Monte Carlo (MCMC), a proposal adds, deletes, or reverses an edge, rejecting moves outside the allowed graph class. Let $q(\G'\mid\G)$ be the proposal probability. The Metropolis--Hastings acceptance probability is
\begin{equation*}
 \min\left\{1,\frac{p(\mathbf X\mid\G')\pi(\G')q(\G\mid\G')}
 {p(\mathbf X\mid\G)\pi(\G)q(\G'\mid\G)}\right\},
\end{equation*}
where $p(\mathbf X\mid\G)$ is the marginal likelihood. When marginalization is unavailable in closed form, an MCMC scheme can instead update parameters jointly with the graph or alternate between graph and parameter updates, using the corresponding acceptance corrections. The \texttt{cyclinbayes} package implements structure MCMC for Bayesian linear non-Gaussian DAG and DCG models \citep{lee2026}; software is available at \url{https://github.com/roblee01/cyclinbayes}.

The sequence of single-edge moves in structure MCMC can have difficulty crossing between modes that differ in many edges. Order MCMC changes the search space by summing over admissible parent sets for each proposed order \citep{friedman2003}. This can alter posterior exploration substantially, but it also changes how a prior is represented: since one DAG can admit several orders, assigning probabilities to orders does not automatically produce a chosen prior on graphs. The computational reformulation must therefore be accompanied by an explicit account of the induced graph prior.

Continuous and sequential graph representations provide further alternatives to local MCMC moves. DiBS \citep{lorch2021} assigns latent embeddings to nodes and uses their inner products to determine edge probabilities, allowing gradients to update a continuous representation of graph uncertainty. A differentiable acyclicity penalty and particle-based variational inference then approximate a graph or joint graph--parameter posterior. DAG-GFlowNet \citep{deleu2022} instead learns to construct a DAG by adding admissible edges and eventually stopping. Its probability-flow objective accounts for the fact that different construction paths can reach the same graph, with the aim of making terminal graph probabilities proportional to an unnormalized posterior score. Thus the two methods approximate posterior sampling through different representations: continuous particles in DiBS and a learned sequential construction in DAG-GFlowNet.

\section{Discrete, categorical, count, and mixed variables}
\label{sec:discrete}

\subsection{Why variable type changes identification}

Moving from continuous to discrete variables changes the restrictions available for identification. Although discrete variables are non-Gaussian, LiNGAM's conclusion relies on a particular linear mixture with independent noises, not simply on the absence of Gaussian distributions. Discrete-data methods must therefore find directional information in other restrictions, such as the form of the mechanism, the complexity of its exogenous noise, or relationships among conditional moments.

Conditional-independence algorithms can use discrete tests and score methods can use multinomial likelihoods, but they generally retain an equivalence-class target. 

\subsection{Entropic and categorical mechanisms}

One route to identifying direction is to restrict the complexity of the exogenous noise. In the original entropic causal inference formulation \citep{kocaoglu2017}, the theoretical argument bounds $H_0$, the logarithm of the number of possible noise values, whereas the practical search approximates a coupling with minimum Shannon entropy $H_1$. 

Compact latent representations of categorical mechanisms offer another way to limit the complexity of an otherwise unrestricted functional representation \citep{cai2018}.

These approaches make explicit why a structural representation alone does not identify direction. If the functions and noise distributions are unrestricted, the same observational joint distribution of $(X,Y)$ can be generated both by a model $Y=f(X,U)$ with $U\indep X$ and by a model $X=g(Y,V)$ with $V\indep Y$. For categorical variables, the first construction draws $X$ from its marginal distribution and uses an independent uniform random variable $U$ to sample a category according to $P(Y=\cdot\mid X)$; reversing the roles gives the second construction. Both reproduce the same contingency table, although their intervention implications can differ. Restrictions on noise complexity or on the form of the mechanism can rule out one of these representations.

For ordinal responses, the ordering of the categories itself supplies additional structure. \citet{ni2022ordinal} use that ordering to restrict the conditional model and establish identifiability. Nominal categories lack this information. Therefore, a different construction is needed: classification with optimal label permutation \citep{ni2022colp} searches over category orders while fitting a parsimonious mechanism. These methods illustrate how the interpretation of the categories determines which restrictions are scientifically meaningful.

Binary variables deserve particular care. Some identifying restrictions that are informative with several categories become saturated or degenerate with two categories. 

\subsection{Conditional moments for count data}

Count models provide a good example of identification through conditional moments. Suppose
\begin{equation*}
 X_j\mid X_{\pa(j)}\sim\operatorname{Poisson}\{\lambda_j(X_{\pa(j)})\}.
\end{equation*}
Then the law of total variance yields
\begin{equation*}
 \Var(X_j)=\E(X_j)+\Var\{\lambda_j(X_{\pa(j)})\}.
\end{equation*}
At a root, the rate is constant and the second term vanishes. A genuinely varying parental rate creates overdispersion after the parents are marginalized. This principle underlies overdispersion scoring for Poisson DAGs \citep{park2015}, which is based on model-specific moment equalities.

To turn this observation into an ordering procedure, suppose a set $S$ of variables has already been ordered and contains only nondescendants of a remaining node $j$. Conditioning on $X_S$ removes the variation explained by those variables while preserving the conditional Poisson argument. The local Markov property and the law of total variance then give
\begin{equation*}
 \Var(X_j\mid X_S)-\E(X_j\mid X_S)
 =\Var\{\lambda_j(X_{\pa(j)})\mid X_S\}\ge0.
\end{equation*}
If $S$ includes all parents of $j$, its rate is fixed after conditioning and the right-hand side is zero. To infer the converse, one needs a nondegeneracy condition: leaving out a parent must leave some variation in the rate. Under this condition, the expected difference $\E\{\Var(X_j\mid X_S)-\E(X_j\mid X_S)\}$ separates nodes whose parents have all been ordered from nodes with a parent still missing.

The restriction to nondescendants explains why this calculation must be used within an ordering procedure. Conditioning on a descendant can change the noise distribution and invalidate the mixture-of-Poissons argument. With the restriction in place, the connection to equal-variance Gaussian discovery is direct: both methods identify an eligible next node by removing variation due to parents that have already been ordered.

The Poisson identity is one instance of a broader relationship between the conditional mean and variance. Quadratic-variance-function DAGs assume that the conditional variance is $\beta_{j0}\mu_j+\beta_{j1}\mu_j^2$, where $\mu_j$ is the conditional mean \citep{park2018}. Their identification result requires $\beta_{j1}>-1$ together with conditions ensuring variation when parents are omitted. This includes several discrete and positive continuous families, but excludes Bernoulli nodes, which lie on the boundary $\beta_{j1}=-1$. Generalized hypergeometric models extend the same reasoning to higher-order moments \citep{park2019ghd}, while sparse Poisson regression and moment-ratio ordering provide high-dimensional estimation procedures \citep{park2019poisson}.

Extra-Poisson variation can also reflect a wrong conditional family, hidden confounding, or the observation process. In those cases the variance contrast need not identify a missing causal parent, motivating sensitivity to alternative count and measurement models.

\subsection{Zero inflation, mixed data, and discrete scores}

In many applications, count datasets contain more zeros than a simple conditional Poisson model can explain. Modeling them explicitly can be important for both fit and identification. \citet{choi2020,choi2023} develop identifiable zero-inflated Poisson DAG models and extensions to zero-inflated generalized hypergeometric distributions under causal sufficiency and nonbinary support. Their parameter-identification results also specify the conditional family and link function. Building on these models, \citet{choi2025} study differences between networks in two populations, combining model-based identification with Bayesian graph learning for zero-inflated counts.

When a dataset contains several variable types, the node-specific mechanisms must fit together into a coherent joint model. Mixed-variable DAGs provide such a construction by combining continuous and discrete conditionals. The restricted linear mixed model of \citet{zeng2022}, for example, establishes identification for its particular joint specification. Its guarantee therefore rests on the way those mechanisms are combined, rather than following from the separate availability of continuous and discrete discovery methods.

Discrete scores offer a further connection to the computational ideas in Section~\ref{sec:computation}. \citet{vo2026} develop a discrete-score criterion that distinguishes sinks under positive conditional probabilities and a condition that conditional randomness does not decrease along arrows. They use it to learn an ordering, paralleling SCORE's use of density derivatives for continuous variables. The operator changes because a mass function has no ordinary derivative with respect to its argument, but the statistical reasoning is similar: a population quantity must distinguish sinks from other nodes, and its sample estimate must preserve that distinction.

\section{Latent confounding and latent structure}
\label{sec:latent}

\subsection{Two different inferential goals}

Allowing unmeasured variables changes both the assumptions and the possible targets of discovery. One aim is to recover causal relations among the observed variables despite confounding; another is to reconstruct the hidden variables and their relationships as well. The first aim can often be pursued with fewer structural restrictions, although it may leave more relationships unresolved. The second requires additional information about how hidden variables influence measurements or about the mechanisms and topology of the latent system. We separate these aims because a method that detects evidence of confounding need not identify the confounders themselves.

In the running example, replace the first two assignments by $X=\lambda_XH+\varepsilon_X$ and $Y=bX+\lambda_YH+\varepsilon_Y$, with $H$ independent of the errors; keep $W=cY+\varepsilon_W$. The separation $X\indep W\mid Y$ survives, but the regression coefficient of $Y$ on $X$ becomes $b+\lambda_X\lambda_Y\Var(H)/\Var(X)$. The causal effect remains $bc$. Thus the observed chain-shaped independence pattern cannot certify the causal interpretation of its fitted coefficients. Gaussian confounding can trade off against $b$ without changing the observed law.

The first aim requires a representation of the causal information that survives marginalization. Simply fitting another DAG to the observed variables need not preserve the original causal interpretation. Ancestral graphs instead accommodate the effects of marginalization and selection \citep{richardson2002}. In conditional-independence discovery, a partial ancestral graph (PAG) summarizes the invariant features of a Markov equivalence class of maximal ancestral graphs. Its endpoint marks record which ancestral statements hold throughout the class and which remain unresolved.

FCI uses conditional-independence tests to search for this information while allowing latent confounders and selection variables \citep{spirtes2000}. The orientation rules of \citet{zhang2008} are sufficient to recover all identifiable endpoint information under the relevant assumptions. RFCI reduces the computational burden by performing fewer tests, with a corresponding loss of some information; it nevertheless has consistency guarantees for the ancestral and independence relationships it represents in sparse high-dimensional settings \citep{colombo2012}. The choice between these procedures thus involves the resolution of the output as well as computational cost.

\subsection{Recovering causal relations among observed variables in the presence of hidden confounding}

Non-Gaussian structural models offer information beyond the observed conditional independences, but the ICA argument changes when hidden common causes are present. A latent-variable extension can be represented schematically as
\begin{equation*}
 X=BX+\Lambda H+\varepsilon,
\end{equation*}
where $H$ contains the latent variables. After solving for $X$, both the structural errors and the hidden variables contribute to the observed mixture. Hence, the basic ICA argument used by LiNGAM no longer directly identifies $B$. \citet{hoyer2008latent} characterize aspects of the resulting ambiguity and the conclusions that non-Gaussian information can still support. Repetitive causal discovery  \citep{maeda2020} takes a more local approach, using residual-based analyses to infer ancestral relationships among observed variables and to indicate relationships affected by confounding.

Beyond linear non-Gaussian models, nonlinearity can interact with assumptions on confounding to restore identification, but not for arbitrary nonlinear latent models. \citet{li2024} study nonlinear DAGs with correlated Gaussian errors and establish generic identification under a sublinear-growth restriction on the mechanisms. This supplies a clear example in which a property of the function class distinguishes causal effects from dependent noises.

Unobserved mediators present another difficulty. Even if direct mechanisms have additive noise, a marginal relationship after integrating out a mediator may not remain in that class. Recent work by \citet{pham2026} develops sufficient conditions for resolving some directions in the presence of unobserved causal paths and backdoor paths, combining residual and conditional-independence information. Therefore, identifiability can be local and path-dependent: some directions may be recoverable even when a complete latent structure is not.

\subsection{Reconstructing latent variables and their causal relationships}

\label{sec:latent-reconstruction}

Reconstructing the latent system requires information about how its variables are expressed in observed measurements. In linear latent-variable models, \citet{silva2006} use measurement structure and constraints on observed associations to recover aspects of that system. Generalized independent noise provides a complementary route: it extends residual-independence reasoning to groups of observed variables and supports discovery of linear non-Gaussian latent graphs \citep{xie2020}. Covariance-rank conditions can also identify latent bottlenecks and hierarchical organization under graphical restrictions \citep{huang2022}. These approaches share the idea that several measurements of a hidden source impose observable constraints beyond ordinary conditional independence.

A simple measurement model illustrates information that conditional independence can miss. Suppose $X_i=\lambda_i H+\varepsilon_i$ for $i=1,2,3,4$, with mutually independent centered noises independent of latent variable $H$, and write $c_{ij}=\Cov(X_i,X_j)$. For $i\neq j$,
\begin{equation*}
 c_{ij}=\lambda_i\lambda_j\Var(H),\qquad
 c_{12}c_{34}-c_{13}c_{24}=0.
\end{equation*}
The second equality is a tetrad constraint. It arises because the cross-covariances all contain products of loadings on the same hidden factor, and hence those products cancel. The observed variables $X_i$ can remain mutually dependent while satisfying the equality, which explains how covariance restrictions can reveal latent measurement structure even when conditional independence supplies little information \citep{silva2006}.

These ideas have both statistical and computational extensions. \citet{ng2024} construct scoring procedures for classes of latent causal models, connecting rank-based identification to exact and continuous score-based search. \citet{chen2024} use tensor-rank conditions for discrete latent structure, where a covariance matrix alone can discard relevant distributional information. The distinction between rank constraints and conditional independences is important: both constrain the observed law, but they exploit different consequences of a latent model.

The meaning of latent recovery also depends on which transformations are left unresolved. Relabeling factors, or rescaling a factor while inversely rescaling its loadings, preserves the graph's adjacency pattern and is usually harmless for structural recovery. A general factor rotation is more consequential: it can change which loadings are zero and thus change the measurement graph. This is why a covariance decomposition alone need not identify latent causal structure. The measurement, rank, and independent-noise restrictions above limit these ambiguities; each theorem must specify the transformations under which its recovered structure or variables remain equivalent.

\subsection{Causal representation learning}
\label{sec:representation}

A related problem is causal representation learning, where the variables in the causal graph must themselves be recovered from high-dimensional measurements. This problem extends latent reconstruction in Section~\ref{sec:latent-reconstruction}: before asking which latent variable causes another, one must determine which latent coordinates the observations distinguish. \citet{scholkopf2021} frame the problem through independent mechanisms and generalization. Different forms of auxiliary information then lead to different identification results. For example, \citet{brehmer2022} use paired observations before and after unknown interventions under weak-supervision assumptions, while CITRIS uses temporal sequences and observed intervention targets \citep{lippe2022}.

Structural restrictions on the measurement graph provide another route. \citet{kong2023hierarchical} study nonlinear latent hierarchical models in which each latent variable has at least two pure children and there is no directed path between siblings. Together with smoothness, information-conservation, and nondegeneracy conditions, these restrictions allow recovery of the hierarchical structure and latent variables up to nodewise invertible transformations. The pure children provide distinct measurements of a hidden source, connecting this nonlinear result to the measurement-model arguments in Section~\ref{sec:latent-reconstruction}. The theorem concerns this restricted hierarchy, rather than arbitrary nonlinear latent graphs.

Environmental variation can instead help separate a latent causal system from an unknown nonlinear observation map. \citet{liu2026weight} study latent linear Gaussian models whose causal coefficients and noise distributions vary with observed context. Their identification result requires sufficiently rich variation and a reference condition under which causal influences vanish; it recovers the latent variables up to permutation and scaling. An earlier polynomial extension in the same line of work, \citet{liu2024polynomial}, permits polynomial latent mechanisms with specified exponential-family noises, again using variation and coefficient restrictions to distinguish latent coordinates. Both works also examine partial identification when some of the required coefficient changes are absent. These results connect representation learning to Section~\ref{sec:heterogeneity}: changes across environments can identify the representation as well as relationships among variables, provided the model restricts which aspects of the mechanisms and measurement map may change.

\section{Directed cyclic graphs and feedback}
\label{sec:cycles}

\subsection{Equilibrium models, solvability, and stability}

Removing the acyclicity restriction from the linear structural model in \eqref{eq:lingam} allows $B$ to represent a directed cyclic graph, with $B_{jj}=0$ and $B_{jk}\ne0$ indicating $k\to j$. The equations $X=BX+\varepsilon$ then have the unique solution $X_*=(I-B)^{-1}\varepsilon$ whenever $I-B$ is invertible. This transformation of the noise distribution defines the observational distribution of the model. Under additional stability conditions, it also describes the limiting distribution of a system adjusting toward equilibrium \citep{mooij2011,bongers2021}.

To see the connection, draw $\varepsilon$ once for each observational unit and hold it fixed during equilibration. Iterating the discrete-time adjustment rule gives
\begin{equation*}
 \begin{aligned}
 X^{(t+1)}&=BX^{(t)}+\varepsilon,\\
 X^{(t)}&=B^tX^{(0)}+\sum_{k=0}^{t-1}B^k\varepsilon\\
 &=B^tX^{(0)}+(I-B^t)(I-B)^{-1}\varepsilon.
 \end{aligned}
\end{equation*}
When the spectral radius $\rho(B)<1$, $B^t\to0$ and the geometric series converges to $(I-B)^{-1}$. Hence $X^{(t)}-X_*=B^t\{X^{(0)}-X_*\}\to0$. For each draw of the noise and any finite initial state, the limit is the structural solution $X_*$. Consequently the convergence is almost sure, and the distribution across observational units converges to the distribution induced by $(I-B)^{-1}\varepsilon$. The spectral-radius condition is necessary and sufficient for convergence to this equilibrium from every initial state; invertibility alone does not ensure convergence.

For continuous time, consider $\dot X(t)=(B-I)X(t)+\varepsilon$ with the same fixed noise. Writing $A=B-I$, variation of constants gives
\begin{equation*}
 \begin{aligned}
 X(t)&=e^{At}X(0)+\int_0^t e^{A(t-s)}\varepsilon\,ds\\
 &=e^{At}X(0)+(I-e^{At})(I-B)^{-1}\varepsilon\\
 &=X_*+e^{At}\{X(0)-X_*\}.
 \end{aligned}
\end{equation*}
The matrix exponential tends to zero exactly when all eigenvalues of $A$ have negative real parts, or equivalently $\max_k\operatorname{Re}\{\lambda_k(B)\}<1$. Under this condition, each trajectory converges to $X_*$ and the distribution across units again converges to the same transformed-noise distribution. This condition differs from $\rho(B)<1$: continuous-time stability bounds the real parts of the eigenvalues, whereas discrete-time stability bounds their moduli. More generally, replacing the differential equation by $\dot X=D\{(B-I)X+\varepsilon\}$ for a positive diagonal matrix $D$ changes adjustment rates without changing the equilibrium; convergence requires the eigenvalues of $D(B-I)$ to have negative real parts. Thus different dynamics can induce the same equilibrium distribution, and equilibrium observations alone need not identify those dynamics.

Holding the noise fixed is essential to the preceding equilibrium interpretation. If a fresh innovation is injected at every step, as in $X^{(t+1)}=BX^{(t)}+\varepsilon^{(t+1)}$ with independent innovations of covariance $\Sigma$, the stable process instead has a stationary covariance satisfying $V=BVB^{\mathsf T}+\Sigma$. This generally differs from the covariance $(I-B)^{-1}\Sigma(I-B)^{-\mathsf T}$ obtained by transforming a single noise. The continuous-time analogue is a stable diffusion $dX=(B-I)X\,dt+L\,dW$, whose stationary covariance satisfies $(B-I)V+V(B-I)^{\mathsf T}+LL^{\mathsf T}=0$. Thus a dynamic interpretation of a cyclic model must specify how noise evolves, not only how the variables affect one another. If the model is also used for interventions, existence and uniqueness must be checked for the intervened systems \citep{bongers2021}.

The factorization implied by a DAG model must also be reconsidered. With dependent endogenous variables linked in a feedback loop, the noise in an assignment need not be independent of its observed parents. Thus, fitting a collection of ordinary regressions and calling their product a likelihood can be incorrect. For a differentiable uniquely invertible transformation from $x$ to noises, a likelihood includes the relevant Jacobian. In the linear independent-noise case,
\begin{equation*}
 p_X(x)=|\det(I-B)|\prod_{j=1}^p
 p_{\varepsilon_j}\{x_j-(Bx)_j\}.
\end{equation*}
For a DAG, a topological ordering makes $I-B$ triangular with unit diagonal, so the determinant equals one. With feedback, that simplification generally fails, and the determinant must be retained in the likelihood. 

\begin{example}
\label{ex:feedback}
For a feedback extension of \eqref{eq:running-chain}, add the return arrow $Y\to X$ while retaining $W=cY+\varepsilon_W$. The first two assignments become
\begin{equation*}
 X=aY+\varepsilon_X,\quad Y=bX+\varepsilon_Y,
 \qquad
 \begin{pmatrix}X\\Y\end{pmatrix}
 =\frac{1}{1-ab}
 \begin{pmatrix}1&a\\b&1\end{pmatrix}
 \begin{pmatrix}\varepsilon_X\\\varepsilon_Y\end{pmatrix}.
\end{equation*}
This example separates solvability, stability, and causal effect interpretation. The observational system has a unique solution when $ab\ne1$, while convergence of the discrete-time adjustment requires the stronger condition $|ab|<1$. Its density is $|1-ab|p_{\varepsilon_X}(x-ay)p_{\varepsilon_Y}(y-bx)$, with the factor $|1-ab|$ accounting for the feedback transformation. Moreover, $\Cov(Y,\varepsilon_X)=b\Var(\varepsilon_X)/(1-ab)$. Hence, the noise in the equation for $X$ is generally correlated with its observed parent $Y$. Regressing $X$ on $Y$ consequently need not estimate $a$; by the same reasoning, $\Cov(X,\varepsilon_Y)=a\Var(\varepsilon_Y)/(1-ab)$ prevents an ordinary regression of $Y$ on $X$ from directly estimating $b$.

A perfect intervention has a different interpretation because it replaces a structural assignment. Fixing $X=x$ removes the first equation and leaves $Y=bx+\varepsilon_Y$, giving $\partial\E\{Y\mid\operatorname{do}(X=x)\}/\partial x=b$. The effect equals that in a one-directional model with the same equation for $Y$: intervening on $X$ breaks the return path through its original assignment. The downstream effect on $W$ therefore remains $bc$, although the feedback loop has changed the observational associations used to estimate it.
\end{example}

\subsection{Identification from conditional independences}

Conditional independences can still provide information about feedback systems, provided the appropriate graphical semantics are used. \citet{richardson1996} developed the cyclic causal discovery algorithm under directed global Markov and faithfulness assumptions, with an output that records invariant features of a cyclic equivalence class. Nonlinear cyclic systems require further care because ordinary d-separation need not describe all of their independence implications. Sigma-separation accounts for strongly connected components, within which every vertex can be reached from every other along directed paths. Using this criterion, \citet{forre2018} develop constraint-based discovery that allows cycles and latent confounders, including data from multiple experimental settings.

This connection to constraint-based discovery extends to FCI. \citet{mooij2020cycles} establish its soundness and completeness properties for suitable simple, sigma-faithful models without selection bias, where simplicity requires unique solvability with respect to every subset of endogenous variables. The resulting PAG records invariant ancestral information under that equivalence theory. It does not specify a unique cyclic adjacency matrix, because conditional independences leave some feedback structures indistinguishable. To resolve additional directions or adjacencies, the methods below introduce distributional restrictions or auxiliary variables.

\subsection{Distributional restrictions and auxiliary variables}
\label{sec:cyclic-distribution}

Non-Gaussianity provides one route beyond cyclic equivalence classes. The LiNG approach of \citet{lacerda2008} applies ICA to identify the mixing matrix up to permutation and scaling, as in LiNGAM. The crucial difference is that a cyclic graph has no topological ordering with which to resolve the permutation ambiguity. Several admissible permutations can therefore produce distinct structural matrices. For systems with disjoint cycles, stability selects the unique candidate within that class; when cycles overlap, further information is needed.

\citet{jin2025long} resolve overlapping cycles through auxiliary instrumental variables. Under causal sufficiency and independent non-Gaussian errors, their sufficient condition requires every directed cycle to contain a node with its own known instrument: an external parent with no direct edge to any other node in the system. An instrumented node shared by several cycles can satisfy the condition for all of them. The known instrument--child relationships rule out alternative ICA permutations. 

Instrumental information can also identify models with Gaussian errors, as illustrated in \citet{sarkar2025mr}. Their model accommodates both feedback and latent confounding. They consider $X=BX+\Gamma Z+\varepsilon$, where $Z$ contains $k$ exogenous instruments, $\varepsilon\sim N(0,\Omega)$, and $I-B$ is invertible. Correlations in $\Omega$ represent dependence among noises that can arise from unmeasured common causes. Under their exogeneity and rank conditions, they establish identifiability when every node has at least one known valid instrument with a nonzero direct effect on that node alone. They further consider invalid instruments that violate the exclusion restriction (i.e., an instrument may directly affect several variables). For this extension, sufficient identification conditions require nonsingular instrument covariance, nonzero $(p-1)$-order minors of the reduced-form matrix $\Pi=(I-B)^{-1}\Gamma$, and the sparsity bound $\max_j\|\Gamma_{j,:}\|_0\le(k-p+1)/2$. The nonzero-minor condition requires sufficiently diverse instrument associations; together with sparse direct instrument effects, it separates the structural matrix $B$ from the observed reduced-form coefficients. Estimation uses second-moment summary statistics and sparse priors on $B$ and $\Omega$.

The comparison with \citet{jin2025long} is therefore a comparison of identifying assumptions. Jin et al. use independent non-Gaussian errors and known instrument--child links that reach every cycle to resolve ICA permutations. Sarkar et al. instead allow correlated Gaussian errors and use instruments to distinguish structural equations. Both exploit external causes, but the instruments supply different information in the two arguments.

Nonlinearity provides another identifying restriction. \citet{mooij2011} study cyclic additive-noise assignments $X=f(X)+\varepsilon$ and establish generic bivariate Gaussian-noise identification under a contraction condition, $\sup_{x,y}|f_X'(y)f_Y'(x)|<1$, excluding affine and specified exceptional mechanism families. Under an invertible noise transformation, the likelihood is
\begin{equation*}
 p_X(x)=|\det\{I-J_f(x)\}|\prod_j p_{\varepsilon_j}\{x_j-f_j(x)\},
\end{equation*}
where $J_f$ is the Jacobian of $f$. Their bivariate identification result supplies a nonlinear observational counterpart to LiNG. Their theorem does not establish identification for general multivariate nonlinear cyclic models.

To separate feedback from latent confounding, \citet{jin2025bayes} combine Gaussian latent factors with independent non-Gaussian structural errors. Their argument proceeds in two stages. Noisy ICA first identifies $(I-B)^{-1}$ up to scaling and permutation, after which stability and disjoint cycles resolve $B$. Transforming the observations to $(I-B)X$ then leaves a factor-analysis problem for the latent loadings. Additional restrictions are needed at both stages: the error families exclude an independent nondegenerate Gaussian component, preventing Gaussian variance from being reassigned freely between factors and errors, and an unordered generalized lower triangular loading structure resolves rotational ambiguity up to column labels and signs. In this loading structure, columns have distinct first-nonzero row positions. Under these restrictions, the Bayesian procedure targets both causal relations among the observed variables and the latent confounding structure.

\citet{dai2026} address a different goal: characterizing distributional equivalence in linear non-Gaussian models with arbitrary latent topology and cycles. Their equivalence notion compares sets of observed distributions generated by graphs. Edge-rank constraints and their relation to ranks of directed-path mixing matrices yield a graphical characterization, class-traversal operations, and a structure-learning procedure for this model class. Recovery of the class of irreducible equivalent models is established with oracle overcomplete ICA and rank-faithfulness assumptions. The result broadens the allowable topology while retaining linearity and distributional restrictions; its target is an equivalence class, rather than the unique restricted model of the preceding approach. Together these developments connect the two goals in Section~\ref{sec:latent}: recovering observed relations and characterizing possible latent structures. Interventions and heterogeneous mechanisms provide further identifying information, reviewed together in Section~\ref{sec:heterogeneity}.

\section{Interventions, multiple environments, and heterogeneous populations}
\label{sec:heterogeneity}

\subsection{Interventional equivalence and experimental design}

The preceding methods extract structure from restrictions on a single observational distribution. Multiple environments offer another source of information, because candidate graphs must now explain how distributions change as well as how variables are associated within one environment. Write these distributions as $P^e$, for environments $e\in\mathcal E$ representing controlled interventions, observed groups, changing external conditions, or latent subpopulations. To use their differences for identification, a model must specify what changes across environments and what is shared. Under known perfect interventions, for example, a mechanism is replaced and its original parental edges are removed. An interventional equivalence class then consists of structures compatible with all the resulting distributions.

In an acyclic causally sufficient model, a perfect randomized intervention on a known target set $I_e$ replaces the corresponding conditional densities by assigned densities $q_j^e$. Its likelihood factorization is
\begin{equation*}
 p^e(x)=\prod_{j\notin I_e}p_j(x_j\mid x_{\pa(j)})
            \prod_{j\in I_e}q_j^e(x_j).
\end{equation*}
The nontarget factors are shared across environments, yielding a pooled likelihood. Candidate directions must explain both observational data and distributions after assignment replacement. Fixed-value interventions have the analogous truncated factorization. Conditioning on an observed value instead retains information about the target's causes \citep{pearl2009}. Soft interventions can alter coefficients or noise distributions while retaining parental influences. These changes imply different likelihoods and cross-environment constraints; specifying the targets and the type of change determines which graphs an experiment can distinguish.

\citet{hauser2012} characterize such classes and develop Greedy Interventional Equivalence Search (GIES) for combining observational and experimental data. \citet{yang2018} extend equivalence theory to more general interventions. The gain in identification depends on which variables are targeted and how interventions affect their mechanisms. For \eqref{eq:running-chain}, randomizing $Y$ independently of its former causes gives $\partial\E\{X\mid\operatorname{do}(Y=y)\}/\partial y=0$ and $\partial\E\{W\mid\operatorname{do}(Y=y)\}/\partial y=c$. Comparing this experiment with the observational law distinguishes the chain from its fork and reverse-chain alternatives under the stated causal model. Repeating an uninformative intervention need not resolve these orientations.

When targets are unknown, identifying the intervention itself becomes part of discovery. Permutation-based procedures address partially or entirely unknown targets \citep{squires2020}, and differentiable causal discovery incorporates intervention information into nonlinear mechanism learning \citep{brouillard2020}. These methods need assumptions that distinguish mechanism changes from observational selection or confounding changes.

Experimental variation can also supply information about cyclic models. \citet{hyttinen2012} use known interventions in linear cyclic systems with latent variables, expressing intervention effects as linear constraints on structural coefficients under sufficient experimental coverage. For nonlinear equilibrium mechanisms, \citet{mooij2013} combine information across experimental conditions through local linear approximations. Once such a model is used to predict an intervention that was not observed, a separate identification question arises: which effect is determined by the available distributions and graphical assumptions? The causal calculus developed for cyclic and latent-variable settings by \citet{forre2020} addresses this step from discovered structure to intervention effects.

These identification results also inform the choice of new experiments. Foundational bounds on the number of interventions needed for causal discovery make the role of experimental coverage explicit \citep{eberhardt2005}.

Scientific validation should target causal predictions. New interventions can discriminate among structures that fit observational data similarly. \citet{agrawal2019} choose experiments to learn a specified graph feature under a budget, using Bayesian design and submodular optimization. \citet{sussex2021} develop approximation guarantees for batches of interventions that can each perturb multiple variables. These approaches connect posterior or equivalence-class uncertainty to the choice of the next experiment.

\subsection{Unknown shifts and invariant mechanisms}

An environment can be informative even when the targets of its intervention are unknown. BACKSHIFT \citep{rothenhausler2015} assumes that environments introduce shifts into a linear system while leaving its structural coefficients and baseline noise covariance unchanged. Comparing covariance matrices across environments then removes the invariant component, including invariant contributions from confounding. What remains can identify the structural coefficients through joint diagonalization, provided the shifts vary sufficiently across environments.

This cancellation can be seen by writing $X^e=BX^e+\varepsilon^e+\delta^e$, with constant $B$, invariant $\Cov(\varepsilon^e)=\Sigma$, shifts $\delta^e$ independent of the noises $\varepsilon^e$, and diagonal shift covariance $D^e$. Let $C^e=\Cov(X^e)$. For a reference environment $e_0$,
\begin{equation*}
 (I-B)(C^e-C^{e_0})(I-B)^{\mathsf T}=D^e-D^{e_0}.
\end{equation*}
The common covariance $\Sigma$ cancels in the difference even when it is nondiagonal. Each environment pair consequently supplies a matrix that the same transformation $I-B$ must diagonalize. Several sufficiently different shift-variance patterns can constrain that transformation, whereas indistinguishable patterns leave ambiguity. This explains both the scope and the limitation of the confounding assumption: invariant confounding is removed by differencing, but changing confounding or changing coefficients would introduce terms that this calculation does not eliminate.

Invariant causal prediction \citep{peters2016} is a local discovery method: it seeks predictor sets for a target variable whose conditional relationship remains stable across environments. When the target mechanism is invariant and the interventions satisfy the required conditions, the procedure can provide error-controlled statements about its causal parents. The resulting set can be conservative when the environments do not perturb enough relevant mechanisms. It does not guarantee the recovery of a complete DAG from a heterogeneous collection of datasets.

For a target $Y$, write $H_{0,S}$ for the hypothesis that a common conditional law $Q_S$ satisfies $P^e(Y\mid X_S)=Q_S(Y\mid X_S)$ in every environment, on the relevant supports. If $\mathcal A_\alpha$ is the collection of sets whose invariance hypotheses are not rejected at level $\alpha$, the basic construction is
\begin{equation}
 \widehat S_{\mathrm{inv}}=\bigcap_{S\in\mathcal A_\alpha}S,
 \qquad
 P\{\widehat S_{\mathrm{inv}}\subseteq\pa(Y)\}\ge1-\alpha.
 \label{eq:invariant-set}
\end{equation}
The coverage statement follows from testing the true parent set validly. If that set is retained among $\mathcal A_\alpha$, then intersecting all retained sets cannot add a variable outside it. The required invariance assumption therefore applies to the target's mechanism, including the absence of interventions on that mechanism \citep{peters2016}. Environments that produce little relevant variation may leave many sets unrejected; taking their intersection can then give a small or empty result while still controlling false inclusion.

Joint Causal Inference offers a broader graphical way to encode environmental changes \citep{mooij2020jci}. It adds context variables describing the environment to the system variables and imposes background assumptions about their relationships, such as exogeneity of the context. Edges from context variables to system variables can then represent mechanisms that change across environments, allowing intervention targets to be learned when they are unknown. This connects the aforementioned shift and invariance arguments: environmental information is represented within an augmented causal model, whose assumptions specify which changes can be used for discovery.

\subsection{Observed heterogeneity in causal effects}

Heterogeneity can be informative even without a shared invariant conditional mechanism. Consider the schematic varying-coefficient model
\begin{equation*}
 X=B(Z)X+\varepsilon,\qquad \varepsilon\mid Z\sim N(0,\Sigma),
\end{equation*}
where $Z$ is an exogenous observed covariate and the error covariance is invariant in $Z$. In this model, the causal effects themselves may vary and an entire conditional distribution need not remain invariant. Instead, the shared noise covariance restricts how alternative causal directions can account for those changes. Under additional identification conditions, this allows variation in $B(Z)$ to distinguish directions that remain ambiguous in a homogeneous linear Gaussian model.

The directional information comes from requiring the noise covariance to remain invariant as effects change. A reverse representation may need changing residual variances to reproduce the same distributions.

A bivariate calculation illustrates the asymmetry. Suppose $X=\varepsilon_X$ and $Y=b(Z)X+\varepsilon_Y$, with independent Gaussian noises independent of $Z$ and constant variances $v_X,v_Y>0$. The reverse conditional Gaussian regression has
\begin{equation*}
 \E(X\mid Y,Z)=\frac{b(Z)v_X}{b(Z)^2v_X+v_Y}Y,
 \qquad
 \Var(X\mid Y,Z)=\frac{v_Xv_Y}{b(Z)^2v_X+v_Y}.
\end{equation*}
If $b(Z)^2$ varies, the reverse residual variance varies, violating the invariant-error-covariance restriction. Applied to the edge $X\to Y$ of Example~\ref{ex:run}, this illustrates how a measured context can identify direction through effect variation; with fixed $c$, its total effect becomes $c\,b(Z)$. Each environment separately remains a directionally ambiguous Gaussian pair, yet their joint variation excludes this reversed representation in the restricted model. It also explains why allowing arbitrary covariance changes would destroy the argument: the reverse model could absorb the apparent asymmetry into its noises.

\citet{zhou2022hetero} establish sufficient identification conditions for three settings: causally insufficient bivariate cyclic graphs; causally insufficient multivariate acyclic graphs; and causally sufficient multivariate cyclic graphs with disjoint cycles. These results do not cover general causally insufficient multivariate cyclic graphs.

\subsection{Latent heterogeneity and individualized graphs}

The variable explaining heterogeneity may itself be unobserved. Bayesian networks with latent trajectory embedding (BN-LTE) represent observations along a low-dimensional latent trajectory and allow graph structure and effects to vary along it \citep{zhou2023individual}. This changes the target from one population graph to a family of observation-specific networks. Smoothness and shared structure allow information to be borrowed across observations so that an individual network is not estimated from a single individual in isolation.

Because the variable indexing heterogeneity is unobserved, identification must determine what can be learned about that variable as well as about the networks it indexes. BN-LTE links heterogeneous effects to identifiable graph structure under causal sufficiency and appropriate noncancellation conditions. With additional conditions, the latent trajectory is determined up to a monotonic transformation, so its numerical scale can remain arbitrary even when its ordering is informative.

The common source of information in these approaches is structured variation, but the structure takes different forms. Invariant prediction keeps a target mechanism fixed while other parts of the system change; varying-coefficient models keep the noise law fixed while effects change. Specifying these shared and varying components is what turns heterogeneity into information about causal structure.

\section{Beyond scalar variables: functional nodes}
\label{sec:functional}

So far, a node has usually represented a scalar measurement. In some applications the scientific variable is instead a whole curve, such as a physiological trajectory, and the target is a relationship between curves rather than between individual measurement times. Functional causal discovery treats each curve as one node, retaining its internal dependence as part of that variable. A function need not be indexed by time, and even when it is, a causal relationship between two functional nodes need not represent a time-lagged effect.

Functional LiNGAM extends non-Gaussian identification to function-valued variables. \citet{yang2022functional} propose functional principal-component coordinates for computation, and \citet{yang2024functional} develop identification for non-Gaussian random vectors and functions in Hilbert spaces. Their central bivariate result requires invertibility of a forward or reverse operator; non-Gaussianity alone does not identify arbitrary functional networks. These approaches connect operator-valued causal assignments to finite-dimensional representations, while accounting for within-function dependence. A Bayesian alternative explicitly models noisy measurements of the curves: \citet{zhou2023functional} develop functional Bayesian networks and a linear non-Gaussian construction with identification in the presence of noisy curve measurements. A schematic functional assignment is
\begin{equation*}
 X_j=\sum_{k\in\pa(j)}\mathcal B_{jk}X_k+U_j,
\end{equation*}
where $X_j$ and $U_j$ belong to function spaces and $\mathcal B_{jk}$ is a suitable linear operator. An edge corresponds to a nonzero functional relationship. For computation, basis expansions translate aspects of the model into relationships among coefficient vectors, while preserving the interpretation of each curve as a node. 

For an integral operator, $(\mathcal B_{jk}X_k)(t)=\int\beta_{jk}(t,s)X_k(s)\,ds$. With finite orthonormal bases $\{\phi_{jr}\}$, coefficient vectors $\xi_j$, and projected noises $u_j$, a working reduction is
\begin{equation*}
 \xi_j=\sum_{k\in\pa(j)}A_{jk}\xi_k+u_j,\qquad
 (A_{jk})_{rs}=\int\!\int\phi_{jr}(t)\beta_{jk}(t,s)\phi_{ks}(s)\,ds\,dt.
\end{equation*}
If the model lies entirely in the retained basis subspaces, this expression is exact; otherwise, truncation leaves a remainder. The distinction matters because the graph concerns relationships between functions. An edge corresponds to a nonzero operator block $A_{jk}$, and hence selecting individual coefficients does not by itself define the functional graph. Under orthogonal basis changes, the block becomes $Q_jA_{jk}Q_k^{\mathsf T}$, preserving whether it is zero and preserving its Frobenius norm. Removing basis directions is different: it can discard the entire component through which one curve affects another. Measurement noise adds another layer between the coefficient vectors and the observed data, which must also be accounted for in estimation.

These representations retain LiNGAM's use of independent non-Gaussian information, but they apply it at the level of functional variables. Coefficients within one curve may be dependent, and the causal relationship is encoded by an operator block rather than a scalar edge weight. Consequently, treating every coefficient as a separate scalar node can change the target graph; functional methods must address the block structure, measurement error, and truncation explicitly.

Feedback between curves brings the functional and cyclic problems together. \citet{roy2023} use a lower-dimensional causal representation of functional observations within a Bayesian directed cyclic model. Their identification theorem combines causal sufficiency and disjoint cycles with spectral conditions on the finite-rank operator, non-Gaussian coefficient noises, a specified noncausal remainder, and a full-column-rank sampling matrix. These assumptions connect what is observed at sampled locations to the latent functional system. In particular, a projection that explains most curve variation need not preserve the components needed to identify causal relations.

Intervening on an entire curve differs from perturbing it at one point. Unless temporal restrictions are imposed on the operators, an edge between curves does not assert that every component of the cause precedes every component of the effect.

\section{Statistical guarantees and uncertainty quantification}
\label{sec:theory}

\subsection{Identification and estimation consistency}

Identification asks whether the population law determines the causal target under the assumed model. Consistency then asks whether an estimator recovers that target as the sample size increases, while uniform consistency controls recovery over a class of laws. The distinction becomes especially important near nonidentifiable configurations. Non-Gaussian noises can approach Gaussian ones, heterogeneous coefficients can approach constants, and nonlinear mechanisms can approach linear ones. Each member of such a sequence may identify its graph, yet the information separating that graph from its competitors can become too weak for reliable recovery at the available sample size.

 For Gaussian conditional-independence methods, strong faithfulness makes this separation quantitative by bounding the relevant nonzero partial correlations away from zero. The geometric analysis of \citet{uhler2013} shows that parameter regions violating strong faithfulness can be substantial. Thus, excluding the measure-zero set of exactly unfaithful parameters does not by itself ensure reliable recovery. A separate difficulty arises when the distributional class is very broad: \citet{shah2020} establish limits on unrestricted conditional-independence testing for continuous distributions and develop the generalized covariance measure under additional restrictions. Together, these results explain why a guarantee for a graphical search must also control the distributional quantities on which its tests depend.

\subsection{Frequentist structure-selection theory}
\label{sec:frequentist-selection}

Frequentist structure-selection guarantees depend on the population information used by the estimator: conditional-independence tests for PC, score comparisons for GES, minimum-trace criteria in Gaussian models, or moment contrasts under non-Gaussianity. The corresponding theory specifies the sparsity, signal, and distributional conditions under which sampling errors do not overturn the identifying comparisons.

Let $T(P)$ be the target identified under the assumed model, such as a CPDAG, the set of compatible causal orderings, or a unique DAG. Selection consistency has the form
\begin{equation*}
 P_0\{\widehat T_n=T(P_0)\}\longrightarrow1;
 \qquad
 \sup_{P\in\mathcal P_n}P\{\widehat T_n\ne T(P)\}\longrightarrow0
\end{equation*}
for pointwise and uniform formulations, respectively. The class $\mathcal P_n$ in a high-dimensional result restricts graph density, minimum signal, and the distributional quantities estimated by the procedure.

The PC analysis of \citet{kalisch2007} illustrates how these requirements enter a high-dimensional result. Bounds on neighborhood size limit the conditional-independence tests needed, while bounds on partial correlations and their separation from zero control the accuracy of those tests. When all relevant decisions are correct, the skeleton and orientation rules recover the target CPDAG. Score-based and hybrid procedures require an analogous analysis of their own comparisons. \citet{nandy2018}, for example, control score differences and the size of graphs encountered during search to establish high-dimensional consistency for GES and adaptively restricted GES.

For penalized least-squares scores, \citet{aragam2019} analyze globally optimal estimators of minimum-trace Gaussian DAG representations. Their support-recovery result uses a minimum-trace identification/gap condition, suitable sparsity and regularization, and a minimum edge signal. 

Under non-Gaussianity, \citet{wang2020} obtain high-dimensional recovery through moment-based ordering information, with bounded in-degree, moment and minimum-signal conditions, and log-concave errors in their high-dimensional analysis. This parallels conditional-independence theory: a population contrast first separates correct and incorrect decisions, and concentration bounds control its empirical error.

\subsection{Bayesian structure-selection theory}
\label{sec:bayesian-selection}

For a graph $\G$ and its parameters $\theta_\G$, the marginal posterior is
\begin{equation*}
 \Pi(\G\mid X^{1:n})\propto\pi(\G)
 \int p(X^{1:n}\mid\theta_\G,\G)\,\pi(d\theta_\G\mid\G).
\end{equation*}
Before studying posterior concentration, one must ensure that prior specification respects the observational equivalences of the model. If two graphs describe the same family of sampling distributions, the likelihood alone cannot distinguish them. Specifying their parameter priors independently can nevertheless produce different marginal likelihoods, because integration weights the common distributions differently. The resulting preference can then be an unintended consequence of parameterization rather than evidence about causal direction.

\citet{geiger2002} provide conditions for coherent parameter priors under which independence-equivalent DAGs have the same marginal likelihood. The key principle is to assign compatible priors to alternative parameterizations of the same observational model. Writing $D$ for the data and $m_{\G}(D)$ for the marginal likelihood, the desired equality is
\[
 m_{\G}(D)=\int p(D\mid\theta_{\G},\G)\,\pi(d\theta_{\G}\mid\G)
          =m_{\G'}(D)
\]
for every dataset $D$ whenever $\G$ and $\G'$ are equivalent in the model under consideration. Their development combines complete-model equivalence and regularity with likelihood modularity, prior modularity, and global parameter independence; in Gaussian models, a normal--Wishart construction supplies compatible priors. Merely giving separate parameter priors with the same distributions or hyperparameter values need not achieve this compatibility.

The equivalence requirement is model-specific, however. DAGs with the same conditional independences need not define the same distribution family once non-Gaussian or other mechanism restrictions are imposed, and hence their marginal likelihoods need not coincide. Posterior consistency must consequently be stated for the equivalence class appropriate to the assumed model.

Recently, \citet{chaudhuri2025} study linear recursive structural models with independent errors that can include both Gaussian and non-Gaussian components. Their theory uses a working Laplace likelihood while permitting true errors from scale mixtures of Gaussians under moment and regularity conditions, with conditional faithfulness required in the general equivalence-class case. Let $\mathcal E(P_0)$ denote the relevant distribution-equivalence class. For fixed dimension $p$, their consistency results establish
\begin{equation*}
 \Pi\{\G\in\mathcal E(P_0)\mid X^{1:n}\}\longrightarrow1
 \quad\text{in }P_0\text{-probability}.
\end{equation*}
The general theorem uses a specifically constructed sample-size-dependent graph prior, for example $\pi_n(\G)\propto\exp\{-n^a|E(\G)|\}$ with $1/2<a<1$. When the identifying assumptions reduce $\mathcal E(P_0)$ to $\{\G_0\}$, concentration on the class becomes selection consistency for the unique DAG. Otherwise, the guarantee concerns the equivalence class as a whole; it neither requires concentration on one representative nor specifies how mass is distributed within the class. 

\subsection{Uncertainty quantification}
\label{sec:uq}

Selection consistency describes eventual recovery, but an analysis of a particular dataset needs a measure of the uncertainty that remains. There are three related questions. First, which directions are unresolved by the population distribution under the model? This is an identification question, summarized by an equivalence class when one is available. Second, how uncertain are we about the structure or class because we observe a finite sample? Confidence sets and posterior distributions address this question in different ways. Third, what do these unresolved structures imply for a scientific target, such as ancestry or an intervention effect? The distinction matters because two graphs may disagree on an edge while agreeing on the effect of interest, and a precisely estimated equivalence class may still leave that effect unidentified.

\subsubsection{Frequentist confidence sets}

\citet{wang2026confidence} address finite-sample structural uncertainty by constructing asymptotically valid confidence sets for causal orderings in additive-error models. Their immediate target is an ordering, from which they derive statements about ancestry and causal effects. For each candidate order, regressions on preceding variables produce residuals whose compatibility with the structural model is assessed using a residual bootstrap. Inverting these goodness-of-fit tests retains the orders that are not rejected. A branch-and-bound algorithm prunes rejected partial orders to reduce computation, although the worst case remains exhaustive. If $p_n(\pi)$ is the test $p$-value for order $\pi$, the confidence set and its pointwise asymptotic guarantee can be written as
\begin{equation*}
 \widehat{\mathcal O}_{\alpha}=\{\pi:p_n(\pi)>\alpha\},\qquad
 \liminf_{n\to\infty}P_0\{\pi_0\in\widehat{\mathcal O}_{\alpha}\}\ge1-\alpha,
\end{equation*}
for a fixed valid ordering $\pi_0$, under the paper's tail, design, and regression-approximation conditions. In the well-specified linear case, one sufficient dimension-growth condition is $p^2\log^{11/2}(n)/n\to0$, with the other constants controlled. Because the guarantee applies to a fixed valid order, it also bounds the probability of retaining at least one valid order. It does not assert simultaneous inclusion of every valid order.

The retained orders can be used to bound ancestral relations and construct effect confidence sets, as discussed below. Further restricting a graph confidence set by selecting parent sets would require additional error control. An empty ordering set can be evidence against the working model, although a valid order can also be rejected with the probability allowed by the test. Multiple environments permit a different, local target: invariant causal prediction controls false inclusions in the parent set of a chosen response, under the invariance and testing assumptions described in Section~\ref{sec:heterogeneity}; see \eqref{eq:invariant-set} \citep{peters2016}.

\citet{strieder2023} instead invert tests of a total effect in causally sufficient Gaussian DAG models with equal error variances. Their likelihood-ratio construction has asymptotic coverage; a split-likelihood alternative has finite-sample coverage. Let $\Theta_{\rm eq}$ be this model class, $\tau(\theta)$ the effect, and $\widehat\theta_1$ a fit using one data split. With likelihood $L_0$ on an independent split, the latter construction is
\begin{equation*}
 R_t=\frac{L_0(\widehat\theta_1)}
 {\sup_{\theta\in\Theta_{\rm eq}:\,\tau(\theta)=t}L_0(\theta)},
 \qquad \mathcal C_\alpha=\{t:R_t\le1/\alpha\}.
\end{equation*}
To see why the split construction controls error, suppose the null value $t$ is the true total effect. The true graph and parameters then belong to the constrained model over which the denominator is maximized, so this denominator is at least as large as the likelihood under the true model. Conditional on the training split, the numerator is the likelihood of a fixed fitted distribution evaluated on independent data. Integrating its ratio to the true likelihood gives at most one, and hence $\E(R_t)\le1$. Markov's inequality therefore bounds the probability of rejecting the true effect value by $\alpha$; retaining the values that are not rejected gives the stated confidence set.

This argument also explains why computation matters for coverage. The denominator must account for all admissible graphs and parameters satisfying the proposed effect value. A numerical local maximum may underestimate that supremum, increasing $R_t$ and potentially rejecting too often. Thus the theoretical guarantee does not automatically extend to an implementation that substitutes an uncertified local fit for the constrained maximum.

The resulting set can be disconnected because uncertainty about direction has consequences different from uncertainty about a coefficient. For data generated by Example \ref{ex:run}, some candidate structures imply no directed path from $X$ to $W$ and hence exactly zero total effect; others imply a nonzero effect whose magnitude must be estimated. The confidence set can accordingly contain an isolated zero together with a range of nonzero values. Within the equal-error-variance model this reflects sampling uncertainty about an identified structure and effect, rather than the population ambiguity of an unrestricted Gaussian DAG model. Compared with the ordering approach, direct effect-test inversion focuses immediately on the scientific target, while relying on stronger distributional restrictions and potentially producing conservative sets.

\subsubsection{Bayesian graph and equivalence-class summaries}
\label{sec:class-uq}

For a structural feature $T(\G)$, its posterior and a credible set satisfy
\begin{equation*}
 \Pi\{T=t\mid X^{1:n}\}=\sum_{\G:T(\G)=t}\Pi(\G\mid X^{1:n}),
 \qquad \Pi\{T\in\mathcal C_\alpha\mid X^{1:n}\}\ge1-\alpha.
\end{equation*}
This construction can summarize an edge indicator, an ancestral relation, or an equivalence-class label, depending on the chosen feature $T$. These feature summaries need not combine into a valid graph: separately thresholding edge probabilities, for example, can produce a cycle even when every posterior draw is acyclic. A representative graph is therefore a separate decision problem that should respect the allowed graph space. The \texttt{cyclinbayes} implementation reports edge, motif, and graph probabilities and selects a representative by posterior expected graph loss \citep{lee2026}. In the instrument-assisted cyclic setting, MR.RGM similarly reports posterior probabilities of network motifs while allowing correlated Gaussian errors \citep{sarkar2025mr}.

For the mixed Gaussian/non-Gaussian model of \citet{chaudhuri2025} reviewed in Section~\ref{sec:bayesian-selection}, define $N=\{j:\varepsilon_j\text{ is non-Gaussian}\}$. A restricted PDAG $\mathcal R(\G;N)$ starts with the ordinary CPDAG of $\G$, restores directions of edges incident to nodes in $N$ according to $\G$, and propagates the resulting compelled orientations. Under the relevant equivalence assumptions, it represents the smaller class compatible with both conditional independences and noise types. If $N$ is empty, the representation reduces to the ordinary CPDAG; if every noise is non-Gaussian, the representation is the uniquely identified DAG.
Their inference procedure summarizes posterior DAG draws through this restricted PDAG representation, pooling mass assigned to equivalent graphs. The authors estimate $N$ separately for each sampled DAG using Bonferroni-adjusted Shapiro--Wilk tests on fitted $L_1$ regression residuals. With retained draws $\G^{(1)},\ldots,\G^{(M)}$, define
\begin{equation*}
 D^{(m)}=\mathcal R\{\G^{(m)};\widehat N(\G^{(m)})\},
 \qquad
 \widehat\Pi_M=\frac1M\sum_{m=1}^{M}\delta_{D^{(m)}},
\end{equation*}
Here $\delta_D$ is a point mass at $D$ and $\widehat\Pi_M$ records the fraction of posterior draws assigned to each restricted PDAG. With correct noise-type information and the applicable equivalence assumptions, equivalent DAGs have the same representation. Pooling their mass thus summarizes uncertainty at the level that the observational model can distinguish. Because the node set $N$ is estimated separately for each draw, the projection itself can also be uncertain, especially when a noise distribution is close to Gaussian.

For these partially directed graphs, the authors define structural Hamming distance $d_H$ by counting unordered node pairs whose status differs: absent, directed either way, or undirected. A representative graph minimizes posterior expected loss over the visited representations $\mathcal V=\{D^{(1)},\ldots,D^{(M)}\}$:
\begin{equation*}
 \widehat D\in\arg\min_{D\in\mathcal V}
       \frac1M\sum_{m=1}^{M}d_H(D,D^{(m)}).
\end{equation*}
This representative is a medoid: it is one of the visited graphs with the smallest average distance to the others. Restricting the minimization to visited representations makes the calculation feasible and ensures that the reported graph occurred in the sample. It also means that the result solves the decision problem over $\mathcal V$, rather than over all possible partially directed graphs.

Following the decision-theoretic credible-ball construction of \citet{wade2018}, originally developed for partitions, let $r_\alpha$ be the smallest nonnegative integer satisfying
\begin{equation*}
 \frac1M\sum_{m=1}^{M}
 \mathbf 1\{d_H(D^{(m)},\widehat D)\le r_\alpha\}\ge1-\alpha,
 \qquad
 \mathcal B_\alpha=\{D:d_H(D,\widehat D)\le r_\alpha\}.
\end{equation*}
The radius expresses posterior dispersion in an interpretable unit: the number of node-pair statuses that differ from the representative. The ball has the stated empirical posterior mass; this construction does not itself establish frequentist coverage. 

Because the same distance can arise from different changes to the network, the radius should be accompanied by the alternative representations and their masses. Together, these summaries show whether uncertainty concerns an edge's presence, its direction, or both.

These summaries separate two questions that a single graph cannot answer. Posterior dispersion across class representations describes uncertainty about which class is supported by the data; undirected edges within a representation record directions that remain unresolved inside that class. A concentrated posterior on one representation can therefore coexist with ambiguity about individual directions.

\subsubsection{Propagation to causal effects}

The scientific consequences of graph uncertainty depend on the target effect. A CPDAG or PAG records unresolved orientations \citep{meek1995,zhang2008}, while a confidence set of orderings can bound ancestral relations \citep{wang2026confidence}. To study effects, one can evaluate the same intervention across compatible graphs, as in IDA \citep{maathuis2009}. In Example~\ref{ex:run}, suppose the errors are Gaussian and the distribution is faithful. The population set of possible \emph{total effects of $X$ on $W$}, over the three Markov-equivalent DAGs, is then $\{0,bc\}$. The forward chain gives the path product $bc$, whereas the fork and reverse chain have no directed path from $X$ to $W$ and give zero. Even unlimited observational data cannot choose between these values within this model class.

Bayesian model averaging propagates graph uncertainty into a common causal estimand $\tau$ through
\begin{equation*}
 \Var(\tau\mid X^{1:n})=
 \E_{\G\mid X^{1:n}}\{\Var(\tau\mid\G,X^{1:n})\}
 +\Var_{\G\mid X^{1:n}}\{\E(\tau\mid\G,X^{1:n})\}.
\end{equation*}

The first term averages parameter uncertainty within each graph, while the second measures variation in the mean effect across graphs. Both contribute when graph choice changes the adjustment or the existence of a causal path. \citet{castelletti2021} use Bayesian model averaging to carry uncertainty about Gaussian graphical structure into intervention-effect inference. Reporting inference conditional on one selected graph omits the second component and can therefore conceal uncertainty that matters for the scientific conclusion.

A frequentist construction uses retained orderings in a different way. For linear SEMs, \citet{wang2026confidence} allocate an error probability of $\alpha/2$ to the ordering confidence set and another $\alpha/2$ to effect intervals obtained from the corresponding valid adjustment sets. Taking the union of these intervals, and including zero when an admissible order puts the outcome before the treatment, accounts for both sources of error. A union bound gives overall asymptotic coverage under their assumptions. As with direct effect-test inversion, the result need not be a single interval, because different structural possibilities can support separated effect values.

\subsection{Comparisons that respect the inferential target}
\label{sec:evaluation}

The distinctions among causal targets should also guide empirical comparisons. A CPDAG estimator should be evaluated against the target CPDAG, since unresolved orientations may be the correct consequence of its assumptions. Similarly, a method allowing latent variables should be assessed on the mixed-graph or ancestral information it aims to recover. Pairwise direction accuracy, causal-order accuracy, graph recovery, and intervention prediction answer different questions; no single metric captures all of them. Structural intervention distance relates certain graph errors to their consequences for intervention predictions \citep{peters2015sid}; its interpretation, however, requires the model conditions under which those predictions are defined and should not be transferred unchanged to arbitrary cyclic or latent systems.

Real-world cause-effect benchmarks provide useful external checks \citep{mooij2016,ni2022colp}. Nevertheless, benchmark ground truth, selection of variable pairs, and the relationship between the data and a method's assumptions require scrutiny. In simulations, the graph generator and coefficient distributions can inadvertently encode a causal order. \citet{reisach2021} show that marginal variance patterns can make common simulated problems unusually favorable to some methods.

For these reasons, simulation studies should vary sample size, sparsity, identification strength, and model misspecification separately. Reporting calibration and failure rates alongside structural accuracy and computational time helps distinguish a method that is uncertain in a difficult problem from one that confidently returns an incorrect graph. Such comparisons are most informative when they explain which assumptions or sources of directional information account for a method's success.

\section{Synthesis and research directions}
\label{sec:discussion}

\subsection{Choosing a method through its assumptions}

Method choice begins by specifying the variables, the observation process, and the causal question. A researcher must then decide which restrictions on confounding, feedback, mechanisms, noises, and environmental changes are scientifically credible. These restrictions determine the information available for identification; tests, scores, ordering procedures, and posterior computation provide ways to extract that information from the data. The computational choice becomes meaningful only in relation to this model and target.

Table~\ref{tab:landscape} brings these considerations together by linking method families to their data types, identifying restrictions, and uncertainty summaries. For example, allowing hidden confounding may justify reporting only certain ancestral relations, whereas credible noise restrictions may identify additional directions. The appropriate resolution of the answer follows from the assumptions that the application can support. It should then be accompanied by an uncertainty statement matched to the reported target.

\begingroup
\spacingset{1.25}
\setlength{\tabcolsep}{3pt}
\renewcommand{\arraystretch}{1.12}
\setlength{\LTcapwidth}{\textwidth}
\begin{longtable}{@{}>{\raggedright\arraybackslash}p{.22\textwidth}>{\raggedright\arraybackslash}p{.13\textwidth}>{\raggedright\arraybackslash}p{.36\textwidth}>{\raggedright\arraybackslash}p{.24\textwidth}@{}}
\caption{Choosing a method by its statistical assumptions. ``Hidden / cycles'' describes the cited model, not an entire research area. Equivalence-class ambiguity, posterior uncertainty, and confidence coverage are different quantities. UQ entries name available interpretations or separate inferential extensions, not guarantees of every algorithm in a row.  }\label{tab:landscape}\\
\toprule
Family and data & Hidden / cycles & Target and key restriction & Uncertainty to report\\
\midrule
\endfirsthead
\multicolumn{4}{l}{\textit{Table \thetable\ continued}}\\
\toprule
Family and data & Hidden / cycles & Target and key restriction & Uncertainty to report\\
\midrule
\endhead
\bottomrule
\endfoot
PC / GES; continuous or categorical with suitable test/score \citep{kalisch2007,chickering2002}
& No / No & CPDAG; Markov and faithfulness or the respective score conditions & Unresolved orientations\\\addlinespace[8pt]
LiNGAM; continuous \citep{shimizu2006}
& No / No & DAG; linearity and independent non-Gaussian errors & Separate ordering-test UQ under its assumptions \citep{wang2026confidence}; see Section~\ref{sec:uq}\\\addlinespace[8pt]
ANM / CAM; continuous \citep{peters2014,buhlmann2014}
& No / No & DAG in restricted additive-noise classes; excludes reversible mechanisms & Ordering confidence sets require additional fit/test conditions; selection alone supplies no coverage\\\addlinespace[8pt]
Equal-variance Gaussian models \citep{petersbuhlmann2014}
& No / No & DAG; common error variance in the specified units & Effect test inversion: asymptotic or split-likelihood coverage \citep{strieder2023}\\\addlinespace[8pt]
Count moments; Poisson / QVF \citep{park2015,park2018}
& No / No & Order and DAG; conditional family, dispersion restriction, and nondegeneracy & Selection consistency differs from calibrated graph or effect confidence\\\addlinespace[8pt]
Categorical / mixed mechanisms \citep{ni2022ordinal,ni2022colp,choi2023,zeng2022}
& No / No & Direction or DAG; ordinal, conditional-family, support, or constructed mixed-model restrictions & Bayesian fits permit posterior summaries\\\addlinespace[8pt]
FCI / RFCI; suitable independence tests \citep{zhang2008,colombo2012}
& Yes / No & PAG or invariant relations; latent-variable Markov and faithfulness conditions & Endpoint ambiguity\\\addlinespace[8pt]
Latent reconstruction; continuous or discrete indicators \citep{silva2006,xie2020,chen2024}
& Yes / No & Latent structure up to specified equivalences; measurement/rank restrictions & State representation ambiguity separately from estimation uncertainty\\\addlinespace[8pt]
LiNG / cyclic IV; continuous \citep{lacerda2008,jin2025long}
& No / Yes & Cyclic structure; non-Gaussian errors plus disjoint-cycle stability or cycle-covering instruments & Candidate structures for LiNG; graph/parameter posteriors in the Bayesian IV model\\\addlinespace[8pt]
Latent cyclic model; continuous \citep{jin2025bayes}
& Yes / Yes & Observed graph and loadings; Gaussian factors, restricted non-Gaussian errors, stable disjoint cycles & Graph and parameter posteriors conditional on loading/noise restrictions\\\addlinespace[8pt]
Known interventions; compatible likelihoods \citep{hauser2012,hyttinen2012}
& Method-specific & Interventional class or causal relations; modeled targets and sufficient experimental coverage & Class ambiguity; confidence requires a separate inferential construction\\\addlinespace[8pt]
Invariant prediction; multiple environments \citep{peters2016}
& No in basic model / No & Target's parents; invariant target mechanism and valid invariance tests & Error-controlled subset of parents, not a full-graph confidence set\\\addlinespace[8pt]
Heterogeneous effects; continuous \citep{zhou2022hetero,zhou2023individual}
& Model-specific & Common or individualized graph; setting-specific confounding/cycle restrictions, structured variation, exogeneity, noncancellation & Bayesian graph/effect summaries; latent-index ambiguity remains distinct\\*\addlinespace[8pt]
Functional nodes; curves \citep{yang2022functional,zhou2023functional,roy2023}
& No  / Yes & Functional DAG or DCG; operator, noise, and observation restrictions & Bayesian functional graph summaries where fitted; basis and measurement uncertainty also matter\\\addlinespace[8pt]
\end{longtable}
\endgroup

\subsection{Open problems}

The first challenge is inference near the boundary of identification. Many results distinguish a generic identifiable case from a special nonidentifiable case, while applications require an assessment of how much directional information a finite sample contains. For instance, the Poisson ordering argument in Section~\ref{sec:discrete} compares conditional variance with conditional mean. When an omitted parent contributes little variation to the rate, the difference $\E\{\Var(X_j\mid X_S)-\E(X_j\mid X_S)\}$ is close to zero even though it is positive in the population. Estimating whether that small difference is zero can require a large sample. Similar difficulties arise with nearly Gaussian noises, nearly constant heterogeneous effects, and mechanisms close to reversible ones. Theory that relates such proximity to sample size would clarify when confidence procedures must retain competing directions and when posterior conclusions are sensitive to the prior.

Second, uncertainty should incorporate the model assumptions that make discovery possible. Current procedures often quantify uncertainty conditional on one functional family, graph class, or noise distribution. A scientifically useful analysis should also expose which conclusions disappear when plausible confounding, measurement error, or mild feedback is introduced. For example, one could bound residual correlations representing unmeasured confounding, or bound the strength of additional feedback edges, and report the ancestral relations or effect signs shared by all compatible models within those bounds. The statistical challenge is to make these sensitivity bounds interpretable and estimable while retaining useful causal conclusions.

Third, the intersection of discrete data, feedback, and latent confounding remains particularly demanding. Conditional distributions cannot simply be multiplied around a cycle, and count or categorical support restrictions interact with equilibrium existence. A theory must specify a coherent data-generating system before adapting a DAG algorithm. The same issue arises when mixed scalar and functional nodes are combined in one feedback network.

A fourth challenge concerns heterogeneous populations, which create both opportunities and ambiguities. Individualized discovery requires borrowing information across observations and distinguishing changing causal structure from changing noise or measurement. Latent embeddings introduce reparameterization ambiguities, and it remains important to understand which individual-level causal statements are invariant to them. Combining observed environmental information with latent heterogeneity may improve identification, but requires assumptions preventing the two explanations from substituting freely for one another.

Finally, computational advances need statistical diagnostics. Continuous optimizers, neural mechanism estimators, and graph-generating samplers make larger analyses feasible, yet their approximation errors can affect the same orientations whose statistical evidence is weakest. Diagnostics for feasibility, search stability, and posterior exploration would make computational performance more reliable. Such diagnostics should accompany checks of the causal model.

The representation-learning results in Section~\ref{sec:representation} raise a related question: which causal conclusions survive uncertainty about the variables themselves? Together with experimental design in Section~\ref{sec:heterogeneity}, these developments suggest studying measurement, discovery, and intervention as connected inferential tasks.

\subsection{Conclusion}

Developments since LiNGAM show how causal discovery can use information beyond conditional independence. Independent non-Gaussian noises can distinguish linear directions; restrictions on nonlinear mechanisms or conditional moments can rule out reverse representations; and interventions or structured heterogeneity can reveal how mechanisms respond to change. Each route supplies a different reason why competing causal structures cannot generate the same data distribution. Allowing latent variables, feedback, or functional nodes requires revisiting that reason, because it changes the distributional model and sometimes the causal object that can be recovered.

For statistical practice, these developments lead to four connected questions: what the assumptions identify, how that target is estimated, whether the computation adequately approximates the estimator, and what uncertainty statement the analysis supports. A discovered graph is most useful when its directions can be traced to defensible assumptions and when the remaining uncertainty is carried through to the scientific relationships or intervention effects of interest.

\par
\interlinepenalty=10000
\bibliographystyle{apalike}
\bibliography{references}
\end{document}